\documentclass[twocolumn,nofootinbib]{revtex4}

\usepackage{graphicx,bm} \usepackage{amsmath,amssymb}
\usepackage{mathrsfs,verbatim,subfigure}

\newcommand{\avg}[1]{\left< #1 \right>} 

\begin{document}

\title{Is $\rho$-Meson Melting Compatible with Chiral Restoration?}

\author{Paul~M.~Hohler}
\email{pmhohler@comp.tamu.edu}
\affiliation{Cyclotron Institute and Department of Physics and Astronomy, Texas A\&M University,
College Station, TX 77843-3366, USA}

\author{Ralf~Rapp}
\email{rapp@comp.tamu.edu}
\affiliation{Cyclotron Institute and Department of Physics and Astronomy, Texas A\&M University,
College Station, TX 77843-3366, USA}

\begin{abstract}
Utilizing in-medium vector spectral functions which describe dilepton data in
ultra-relativistic heavy-ion collisions, we conduct a comprehensive evaluation of
QCD and Weinberg sum rules at finite temperature. The starting point is our recent
study in vacuum, where the sum rules have been quantitatively satisfied using
phenomenological axial-/vector spectral functions which describe hadronic
$\tau$-decay data. In the medium, the temperature dependence of condensates and
chiral order parameters is taken from thermal lattice QCD where available,
and otherwise is estimated from a hadron resonance gas. Since little is known
about the in-medium axialvector spectral function, we model it with a
Breit-Wigner ansatz allowing for smooth temperature variations of its width
and mass parameters. Our study thus amounts to testing the compatibility of
the $\rho$-broadening found in dilepton experiments with (the approach toward)
chiral restoration, and thereby searching for viable in-medium axialvector
spectral functions.
\end{abstract}


\maketitle

\section{Introduction}
\label{sec:intro}
The structure of the QCD ground state is reflected in its observable hadron spectrum.
In vacuum, the formation of quark and gluon condensates leads to the generation
of hadron masses and the spontaneous breaking of chiral symmetry (SBCS). The latter
induces mass splittings of ca.~0.5\,GeV for chiral partners in the light-hadron
spectrum, {\it e.g.}, between $\pi$-$\sigma$ or $\rho$-$a_1$. In a hot medium, chiral
symmetry is restored across a region around a pseudo-critical temperature of
$T_{\rm pc}$$\simeq$160\,MeV~\cite{Borsanyi:2010bp,Bazavov:2011nk}. A long-standing
question is how this restoration manifests itself in the hadron spectrum, {\it i.e.},
what its observable consequences are. Dilepton data from ultra-relativistic
heavy-ion collisions (URHICs)~\cite{Arnaldi:2008fw,Adamova:2006nu,Geurts:2012rv} are
now providing strong evidence that the $\rho$ resonance ``melts" when the system
passes through the pseudo-critical region~\cite{Rapp:2013ema}, while experimental
access to the in-medium $a_1$ spectral functions (e.g., via $a_1\to\pi\gamma$) remains
elusive. Thus, to test whether the $\rho$ melting in the vector channel signals chiral
restoration, a theoretical evaluation of the in-medium axialvector spectral function
is needed.

A straightforward approach to calculate the in-medium axialvector spectral
function, by using a chiral Lagrangian paralleling the treatment of the $\rho$ meson,
turns out to be challenging~\cite{Urban:2001uv}. For example, the widely used scheme
of implementing the $\rho$ and $a_1$ mesons into the pion Lagrangian through a local
gauging procedure causes considerable problems in describing the vacuum spectral
functions as measured in hadronic $\tau$ decays~\cite{Barate:1998uf,Ackerstaff:1998yj},
which led some groups to abandon
the local gauging procedure~\cite{Urban:2001ru,Parganlija:2010fz}. In the present
work, we adopt a more modest approach to this problem, by utilizing in-medium sum
rules. Specifically, we adopt the well-known Weinberg sum rules
(WSRs)~\cite{Das:1967ek,Weinberg:1967kj,Kapusta:1993hq} which relate (moments of) the
difference between vector and axialvector spectral functions to operators signifying
SBCS. Using available calculations of the in-medium $\rho$ spectral function together
with temperature dependent order parameters as an input, we ask whether {\em a} (not
necessarily {\em the}) axialvector spectral function can be found to satisfy the in-medium sum
rules. To tighten our constraints, we simultaneously employ
finite-temperature QCD sum rules (QCDSRs)~\cite{Shifman:1978bx,Shifman:1978by}
in vector and axialvector channels, which additionally involve chirally
invariant condensates. Related works have been carried out, e.g.,  in the low-temperature
limit~\cite{Marco:2001dh,Holt:2012wr}, for heavy-quark channels~\cite{Hilger:2011cq},
or focusing on chirally odd condensates in the vector channel only~\cite{Hilger:2010cn}.

The present analysis builds on our previous work~\cite{Hohler:2012xd} where QCD and
Weinberg sum rules have been tested in vacuum with vector and axialvector spectral
functions that accurately fit hadronic $\tau$-decays.
The combination of four WSRs turned out be a rather sensitive probe of the spectral
functions,  allowing, {\it e.g.}, to deduce the presence of an excited axialvector
meson, $a_1'$. This makes for a promising tool at finite temperature ($T$), aided by
an experimentally tested in-medium vector spectral function and
in-medium condensates from lattice QCD (lQCD). In the absence of reliable microscopic
models for the $a_1$ and the excited states, the price to pay is the {\it a priori}
unknown in-medium behavior of these states. However, with  guidance from
model-independent chiral mixing theorems to constrain the $T$ dependence of the
higher states, one can still hope for a sensitive test of the in-medium $a_1$ spectral
function, and to gain novel insights into (the approach to) chiral restoration
in the $IJ^P=11^\pm$ chiral multiplet. This is the main objective of our work.

The Letter is organized as follows. We recall the in-medium QCDSRs and WSRs
in Sec.~\ref{sec:sumrule} and specify the $T$ dependence of their ``right-hand
sides" (condensates) in Sec.~\ref{sec:cond}. The  finite-$T$ axial-/vector spectral
functions (``left-hand sides")
are detailed in Sec.~\ref{sec:FTspec}, followed by quantitative sum rule
analyses in Sec.~\ref{sec:results}. We conclude in Sec.~\ref{sec:conc}.

\section{Finite Temperature Sum Rules}
\label{sec:sumrule}
The basic quantity figuring into WSRs and QCDSRs is the isovector
current-current correlator in the vector ($V$) and  axialvector ($A$) channels,
\begin{equation}
\Pi_{V,A}^{\mu\nu}(q^2) = - i \int d^4 x \ e^{i x q}
\avg{T \vec{J}_{V,A}^{\mu}(x) \vec{J}_{V,A}^{\nu}(0) } \ .
\end{equation}
In the quark basis with two light flavors, the currents read
$\vec{J}_V^\mu = \bar{q} \vec{\tau} \gamma^\mu q$ and
$\vec{J}_A^\mu = \bar{q} \vec{\tau} \gamma^\mu \gamma_5 q$,
($\vec{\tau}$: isospin Pauli matrices). From here on, we focus on charge-neutral states
(isospin $I_3$=0) and drop isospin indices. In vacuum,
the currents can be decomposed into 4D transverse and longitudinal components as
\begin{equation}
\Pi_{V,A}^{\mu\nu}(q^2) = \Pi_{V,A}^T(q^2) \left(-g^{\mu\nu} + \frac{q^\mu q^\nu}{q^2}\right)
+ \Pi_{V,A}^L(q^2) \frac{q^\mu q^\nu}{q^2} \ .
\end{equation}
Vector-current conservation implies $\Pi_V^L(q^2)$=0, while the pion pole induces
the partial conservation of the axialvector current (PCAC),
\begin{equation}
\Pi_A^L (q^2) = f_\pi^2 q^2 \delta(q^2-m_\pi^2) \  .
\end{equation}
Lorentz symmetry breaking at finite $T$
splits the 4D-transverse polarization functions
into 3D-transverse and 3D-longitudinal parts. From here on, we focus on
vanishing 3-momentum ($\vec{q}$=0), for which the 3D components are degenerate.
We define pertinent spectral functions as
\begin{equation}
\rho_{V,A} = -\frac{{\rm Im}\Pi_{V,A}^{T}}{\pi} \ , \
\rho_{\bar A}= \rho_{A} - \frac{{\rm Im}\Pi_{A}^{L}}{\pi} \ .
\end{equation}

The QCDSRs equate a dispersion integral on the left-hand-side (LHS) to an operator
product expansion (OPE) on the right-hand-side (RHS); for the axial-/vector channels
they read~\cite{Hatsuda:1992bv,Leupold:1998bt,Zschocke:2002mn}
\begin{eqnarray}
&&\!\!\!\!\frac{1}{M^2}\!\int_0^\infty \!ds \frac{\rho_{V,\bar{A}}(s)}{s} e^{-s/M^2}
  =  \frac{1}{8\pi^2} \left(1+\frac{\alpha_s}{\pi}\right)
+\frac{m_q \langle\bar{q}q\rangle}{M^4} \nonumber\\&
&\!\!\!\!+\frac{1}{24 M^4}\langle\frac{\alpha_s}{\pi} G_{\mu\nu}^2\rangle
-  \frac{\pi \alpha_s}{M^6} \frac{(56,-88)}{81}  \langle \mathcal{O}_4^{V,A} \rangle
 \\
& &\!\!\!\!+\sum_h \frac{\langle \mathcal{O}^{d=4,\tau=2}_h \rangle_T}{M^4}+\frac{\langle\mathcal{O}^{d=6,\tau=2}_h
\rangle_T}{M^6}+\frac{\langle \mathcal{O}^{d=6, \tau=4}_h \rangle_T}{M^6} \ldots \ , \nonumber
\end{eqnarray}
where the space-like $q^2$ is traded for the Borel mass $M^2$ by a
standard Borel transform. On the RHS, we include all operators up to dimension-6,
i.e., the common scalar operators already present in the vacuum (quark,
gluon, and 4-quark condensates, $\avg{\bar{q}q}$,
$\avg{\frac{\alpha_s}{\pi} G^2_{\mu\nu}}$, and $\langle\mathcal{O}_4^{V,A}\rangle$,
respectively), as well as non-scalar operators induced by thermal hadrons ($h$),
organized by dimension ($d$) and twist ($\tau$).
The $T$ dependencies are detailed in Sec.~\ref{sec:cond}.

The WSRs relate moments of the difference between the vector and axialvector spectral
functions to chiral order parameters. Their formulation at finite $T$ was first carried
out in Ref.~\cite{Kapusta:1993hq}. Subtracting the two channels of the finite-$T$ QCDSRs
from one another, Taylor-expanding the Borel exponential, and equating powers of $M^2$
on each side of the sum rule yields
\begin{eqnarray}
({\rm WSR}\, 1)& \quad  \int_0^\infty \!ds \,
\frac{\Delta\rho(s)}{s} = f_\pi^2 \ ,
\label{eq:WSR1} \\
({\rm WSR}\, 2)& \quad \int_0^\infty\! ds\, \Delta\rho(s)
=   f_\pi^2 m_\pi^2 = -2 m_q \langle \bar{q}q \rangle \ ,
\label{eq:WSR2}\\
({\rm WSR}\, 3)& \quad \int_0^\infty ds s \Delta\rho(s)
= - 2 \pi \alpha_s \langle \mathcal{O}_4^{SB} \rangle \ ,\label{eq:WSR3}
\end{eqnarray}
where $\Delta \rho = \rho_V - \rho_A$. The chiral breaking 4-quark condensate is
given by the axial-/vector ones as
\begin{equation} \label{eq:q4sbdef}
\avg{\mathcal{O}_4^{SB}} = \frac{16}{9}\left( \frac{7}{18} \avg{\mathcal{O}_4^V} + \frac{11}{18} \avg{\mathcal{O}_4^A}\right) \, .
\end{equation}
Since the WSRs only contain chiral order parameters, they are particularly
sensitive to chiral symmetry restoration, whereas the QCDSRs are channel specific
thus providing independent information.

\section{In-Medium Condensates}
\label{sec:cond}
We now turn to the $T$ dependence of each condensate figuring into the QCDSRs.
To leading order in the density of a hadron $h$ in the heat bath, the in-medium
condensate associated with a given operator $\mathcal{O}$ can be approximated by
\begin{equation}
\label{eq:opT}
\langle \mathcal{O} \rangle_T \simeq \langle \mathcal{O}\rangle_0 + d_h
\int \frac{d^3 k}{\left(2 \pi\right)^3 2 E_h}
\langle h(\vec{k})|\mathcal{O}|h(\vec{k})\rangle n_h(E_h) \ ,
\end{equation}
where $\langle \mathcal{O}\rangle_0$ is the vacuum value of the operator,
$\langle h(\vec{k})|\mathcal{O}|h(\vec{k})\rangle$ its hadronic matrix element,
$E_h^2$=$m_h^2+\vec{k}^2$, and $d_h$, $m_h$, and $n_h$ are the hadron's spin-isospin
degeneracy, mass, and thermal distribution function (Bose ($n_b$) or Fermi ($n_f$)),
respectively.
Working at zero baryon chemical potential ($\mu_B$=0), we absorb anti-baryons
into the degeneracy factor of baryons. Corrections to Eq.~(\ref{eq:opT}) figure via
multi-hadron matrix elements of the operator.

We approximate the medium by a hadron resonance gas (HRG) including all confirmed states
with mass $m_h$$\leq$~2\,GeV~\cite{pdg}. For the temperatures of interest here,
$T$$\lesssim$\,170\,MeV, the HRG is known to reproduce the equation of state from lQCD
quite well~\cite{Karsch:2003vd}.
Since the calculation of the in-medium $\rho$ spectral function is also based on HRG
degrees of freedom, the OPE and spectral function sides of the sum rules are evaluated
in the same basis. For the subsequent discussion, we define the integrals
\begin{equation}
I_n^h = d_h \int \frac{d^3 k}{(2\pi)^3 E_h} k^{2n-2} n_{h}(E_h) \ .
\end{equation}
Note that $m_h I_1^h$ is the scalar density, $\varrho_s^h$.

\subsection{Quark Condensate}
The HRG correction to the quark condensate is~\cite{Gerber:1988tt,Leupold:2006ih}
\begin{equation} \label{eq:q2T}
\begin{split}
&\frac{\avg{\bar{q}q}_T}{\avg{\bar{q}q}_0} = 1 - \frac{\varrho_s^\pi}{2 m_\pi f_\pi^2} -
\frac{\varrho_s^K}{4 m_K f_K^2} - \frac{\varrho_s^\eta}{6 m_\eta f_\eta^2}
- \frac{\varrho_s^{\eta'}}{ 3 m_{\eta'} f_{\eta'}^2} \\
& \qquad \quad \quad - \sum_B \frac{\sigma_B}{f_\pi^2 m_\pi^2} \varrho_s^B-
\sum_M \frac{\sigma_M}{f_\pi^2 m_\pi^2} \varrho_s^M- \alpha T^{10} \ .
\end{split}
\end{equation}
The Goldstone boson contribution can be inferred from current algebra (with decay constants
given in Tab.~\ref{tab:parq2T}). The contributions from baryons ($B$) and other mesons ($M$)
can be derived from the HRG partition function via $\partial \ln Z/\partial m_q$, which is
nothing but the in-medium condensate.  They are determined by their $\sigma$-terms which to
lowest order are given by the (current) quark masses, $m_q$, of the light valence quarks in
the hadron~\cite{Gasser:1990ce}. However, important contributions arise from the
hadron's pion cloud~\cite{Jameson:1992ep,Birse:1992he}.
We write
\begin{equation}
\sigma_h = \sigma_q^{\rm bare} + \sigma_\pi^{\rm cloud}  \equiv \sigma_0\, m_q\, (N_q-N_s)
\label{sigh}
\end{equation}
where $N_q$ ($N_s$) is the number of all (strange) valence quarks in $h$.
We adjust the proportionality constant to $\sigma_0$=2.81, to recover the recent value,
$\sigma_N$=59\,MeV~\cite{MartinCamalich:2010fp}, of the nucleon and assume it to be
universal for all hadrons. This leads to fair agreement with estimates of $\sigma_h$
for other ground-state baryons~\cite{MartinCamalich:2010fp}.
Note that the decomposition of the $\sigma$ terms into quark core and pion cloud effects
parallels the medium effects of the $\rho$ spectral function~\cite{Rapp:2012zq}.

\begin{figure}[t!]
\centering
\includegraphics[width=.42\textwidth]{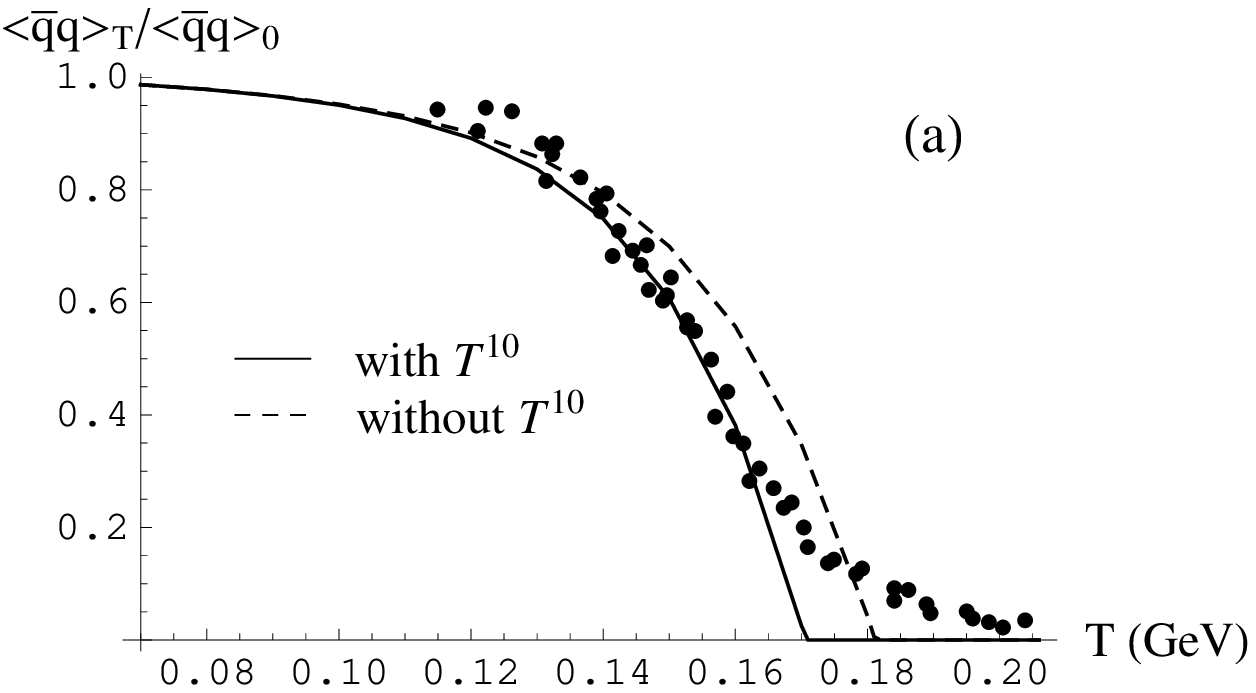}
\includegraphics[width=.42\textwidth]{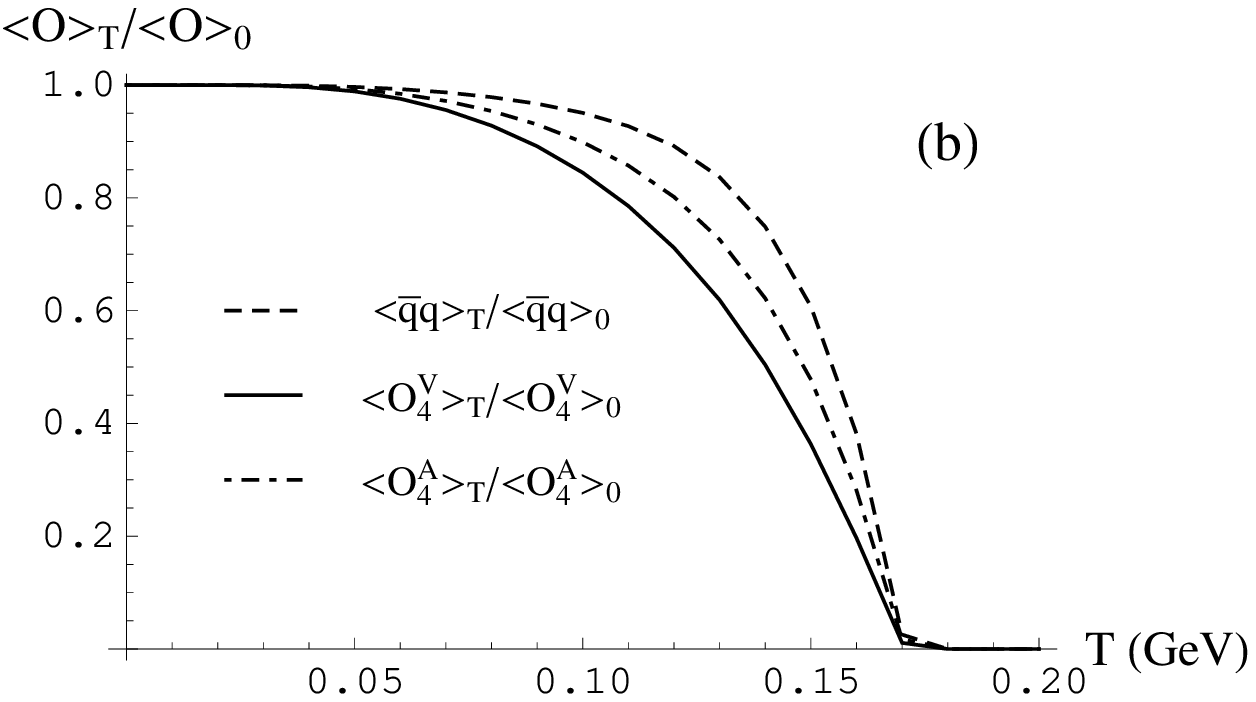}
\caption{Temperature dependence of: (a) the quark condensate relative to its vacuum
value, compared to thermal lQCD data~\cite{Borsanyi:2010bp};
(b) axial-/vector 4-quark condensates relative
to their vacuum values, compared to the quark condensate.}
\label{fig:cond}
\end{figure}

Our HRG results reproduce lQCD ``data"~\cite{Borsanyi:2010bp} for
$T$$\lesssim$140\,MeV, see Fig.~\ref{fig:cond}(a).
To improve the agreement at higher $T$ without affecting the low-$T$ behavior, we
introduced a term $\alpha T^{10}$ on the RHS of Eq.~(\ref{eq:q2T}),
with $\alpha$=1.597~$\cdot 10^7$\,GeV$^{-10}$.
The quark condensate then vanishes slightly above $T$=170\,MeV, signaling
the breakdown of our approach.
Choosing a somewhat higher power in $T$ (with accordingly adjusted $\alpha$) has no
significant impact on our results, while a smaller power adversely affects the
agreement with lQCD data at low $T$.

\begin{table}[!b]
\begin{center}
\begin{tabular}{|c|cccccc|}
\hline
Parameter & $f_\pi$  & $f_K$ &$f_\eta$ & $f_{\eta'}$ & $m_q$ & $m_\pi$ \\
\hline
Value (MeV) & 92.4 & 113 &124 & 107 & 7 & 139.6 \\
\hline
\end{tabular}
\end{center}
\vspace{-0.3cm}
\caption{Numerical values of key parameters figuring into Eq.~(\ref{eq:q2T}).
For hadron masses not listed we take averages from the particle data
group~\cite{pdg}.}
\label{tab:parq2T}
\end{table}
\subsection{Gluon Condensate}
For the gluon condensate, the contributions from pions and nucleons have been
evaluated in Refs.~\cite{Cohen:1991nk,Hatsuda:1992bv,Zschocke:2002mn}. The
HRG effect can be inferred from the trace anomaly,
\begin{equation}
\theta^\mu_\mu =  -\frac{9}{8} \frac{\alpha_s}{\pi}G^2_{\mu\nu} + \sum\limits_{q} m_q \bar qq \ ,
\end{equation}
by calculating  $\Delta \! \avg{\theta^\mu_\mu} = \epsilon -3P = \sum_h m_h \varrho_s^h$ to obtain
\begin{equation}
\!\!\Delta\!\avg{\frac{\alpha_s}{\pi}G^2_{\mu\nu}} = -\frac{8}{9} \left[ \Delta \! \avg{\theta^\mu_\mu}-2 m_q \Delta\!\avg{\bar{q}q} - m_s \Delta\!\avg{\bar{s}s}\right]  .
\end{equation}
The change in light-quark condensate is taken from Eq.~(\ref{eq:q2T}).
For the strange-quark condensate, we assume its suppression from individual resonances
to scale with the valence strange-quark content of each hadron $h$, paralleling the procedure
of determining the $\sigma$-term for each hadron. One has
\begin{equation}
m_s \Delta\!\avg{\bar{s}s} = \sum_h \frac{N_s}{N_q-N_s} \left(2 m_q \Delta\!\avg{\bar{q}q}_h \right),
\end{equation}
where $\Delta\!\avg{\bar{q}q}_h$ is from Eq.~(\ref{eq:q2T}).
The HRG suppression of the gluon condensate reaches 13\% at $T$=170\,MeV.

\subsection{Four-Quark Condensates}
For medium dependence of the vector and axialvector 4-quark condensates
induced by Goldstone bosons, we adopt the results from current
algebra~\cite{Hatsuda:1992bv}. For the non-Goldstone bosons and baryons,
arguments based on the large-$N_c$ limit~\cite{Leupold:2005eq,Leupold:2006ih}
suggest a factorization approximation, i.e., the medium effect linear in their
(scalar) density amounts to a factor of 2 times the reduction in the quark condensate,
with the same factorization parameter as in vacuum (we have checked that an increase
of the in-medium factorization parameter by a factor of 2 has a negligible impact
on the  OPEs and thus on the resulting spectral functions).
The $T$ dependence of the vector and axialvector 4-quark condensates then
takes the form
\begin{eqnarray} \label{eq:V4qFT}
&&\frac{\langle \mathcal{O}_4^{V,A} \rangle_T}{\langle \mathcal{O}_4^{V,A} \rangle_0} =
1 - \frac{(12/7, 12/11)}{m_\pi f_\pi^2} \varrho_s^\pi
- \frac{(9/14, 9/22)}{m_K f_K^2} \varrho_s^K \nonumber\\
&&\quad \ \   - \sum_B \frac{2\sigma_B}{f_\pi^2 m_\pi^2} \varrho_s^B
-\sum_M \frac{2\sigma_M}{f_\pi^2 m_\pi^2} \varrho_s^M + \beta_{V,A} T^{10} \ .
\end{eqnarray}

As for the quark condensate, we augmented the $T$ dependence by a term
$\beta_{V,A} T^{10}$. Since thermal lQCD data are not available for 4-quark condensates, we
adjusted $\beta_{V,A}$ for each channel to render them vanishing at the same temperature as
the quark condensate, resulting in $\beta_V$=$3.05 \cdot 10^7 {\rm GeV}^{-10}$ and
$\beta_A$=$1.74 \cdot 10^7 {\rm GeV}^{-10}$.
The $T$ dependence of the chiral breaking 4-quark condensate follows from the axial-/vector ones
via Eq.~(\ref{eq:q4sbdef}); relative to the quark condensate, their initial fall-off is faster but
slows down above $T$$\simeq$140\,MeV, cf.~Fig.~\ref{fig:cond}(b).

\subsection{Non-Scalar Condensates}
Hadrons in the heat bath also induce non-scalar condensates. For our
QCDSR analysis the relevant ones are of dimension-4 twist-2,
$\avg{\mathcal{O}^{d=4,\tau=2}}_T$, dimension-6 twist-2,
$\avg{\mathcal{O}^{d=6,\tau=2}}_T$, and dimension-6 twist-4,
$\avg{\mathcal{O}^{d=6,\tau=4}}_T$. We adopt their $T$ dependence as elaborated
in Refs.~\cite{Hatsuda:1992bv,Leupold:1998bt,Zschocke:2002mn}, given by each hadron as
\begin{eqnarray}
\langle \mathcal{O}^{d=4, \tau=2}_h \rangle_T &=& \frac{A_2^h}{4}\left(m_h^2 I_1^h+\frac{4}{3} I_2^h\right),\nonumber\\
\langle \mathcal{O}^{d=6, \tau=2}_h \rangle_T &=& -\frac{5 A_4^h}{24}\left(m_h^4 I_1^h + 4 m_h^2 I_2^h +\frac{16}{5} I_3^h\right),\nonumber\\
\langle \mathcal{O}^{d=6, \tau=4}_h \rangle_T &=& \frac{B_2^h}{4}\left(m_h^2 I_1^h+\frac{4}{3} I_2^h\right).
\end{eqnarray}

The parameters $A_2$ and $A_4$,
which control the twist-2 operators, are related to moments of parton distribution functions
for the $u$ and $d$ quarks in the hadron
\begin{equation} \label{eq:an}
A_n = 2 \int_0^1 dx x^{n-1} (\bar{q}(x)+q(x)) \ .
\end{equation}
One can think of $A_2$ as twice the momentum fraction of the up and down
quarks in the hadron, with $A_4$ a higher moment.
Their values are reasonably well known for the pion and nucleon, $A_2^\pi = 0.97$,
$A_4^{\pi}$=0.255, $A_2^N$=1.12, $A_4^N$=0.12, while there is substantial uncertainty
for other hadrons. For baryons, we assume $A_2$ and $A_4$ to be identical
to the nucleon values, but weighted by the light-quark fraction; {\it e.g.},
the $A_2$ of the $\Lambda$ is  $\frac{2}{3}A_2^N$ . The kaons and etas are
approximated with the pion's parton distribution functions, reduced by the strange-quark
content.
For other mesons, Eq.~(\ref{eq:an}) is used with the nucleon parton
distributions functions, rescaled by the valence-quark content and also reduced by
the strange-quark content. This gives $A_2$=0.801 and $A_4$=0.086 for non-strange
mesons.  The $B_2$'s are related to integrals of the twist-4 part of the spin-averaged
(longitudinal) structure function, $F_{2(L)}^{\tau=4}$~\cite{Choi:1993cu,Leupold:1998bt}.
For the nucleon,
it has been extracted as $B_2^N$=$-$0.247\,GeV$^2$. Since there is no empirical
information for other hadrons, we assume their $B_2$ to be the same as for the
nucleon (suppressed by the strange-quark content); varying it by a factor
of 2 produces no noticeable changes in the final spectral functions.
Gluonic contributions are believed to be numerically
insignificant~\cite{Hatsuda:1992bv,Leupold:1998bt} and have been neglected.

\section{Finite Temperature Spectral Functions}
\label{sec:FTspec}
\begin{figure}[tb]
\centering
\includegraphics[width=.4\textwidth]{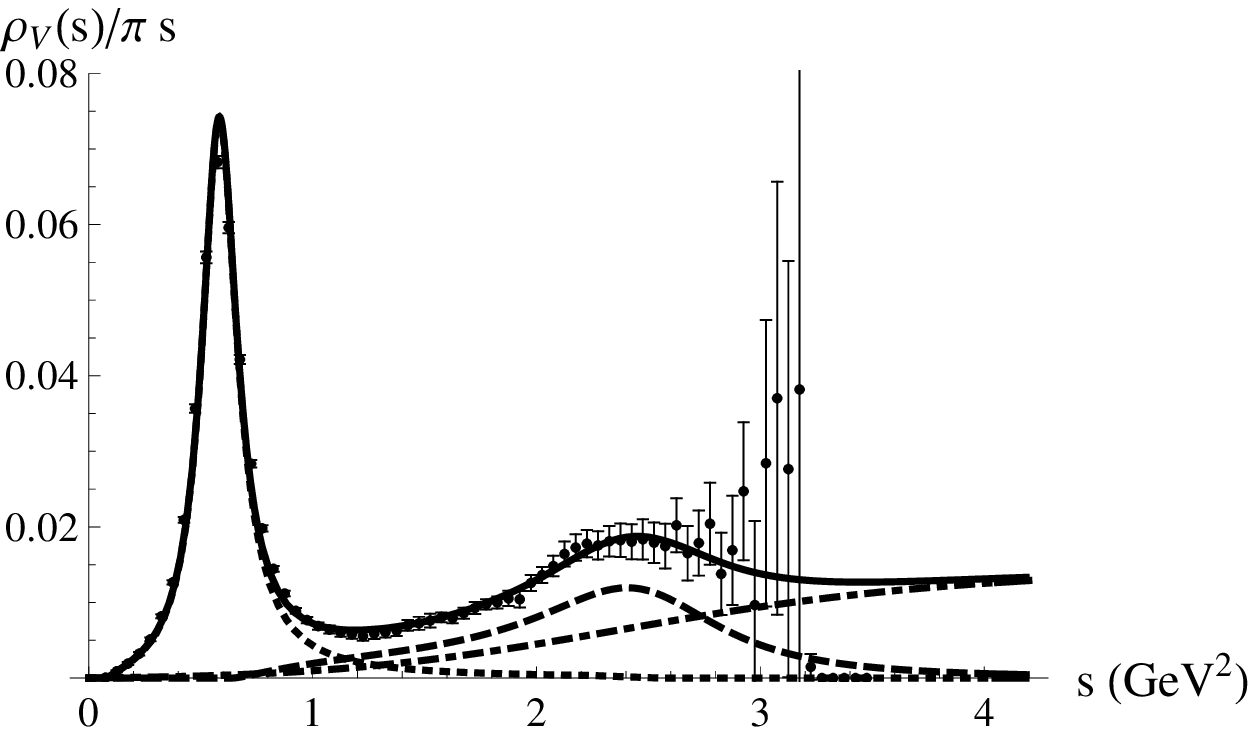}
\includegraphics[width=.4\textwidth]{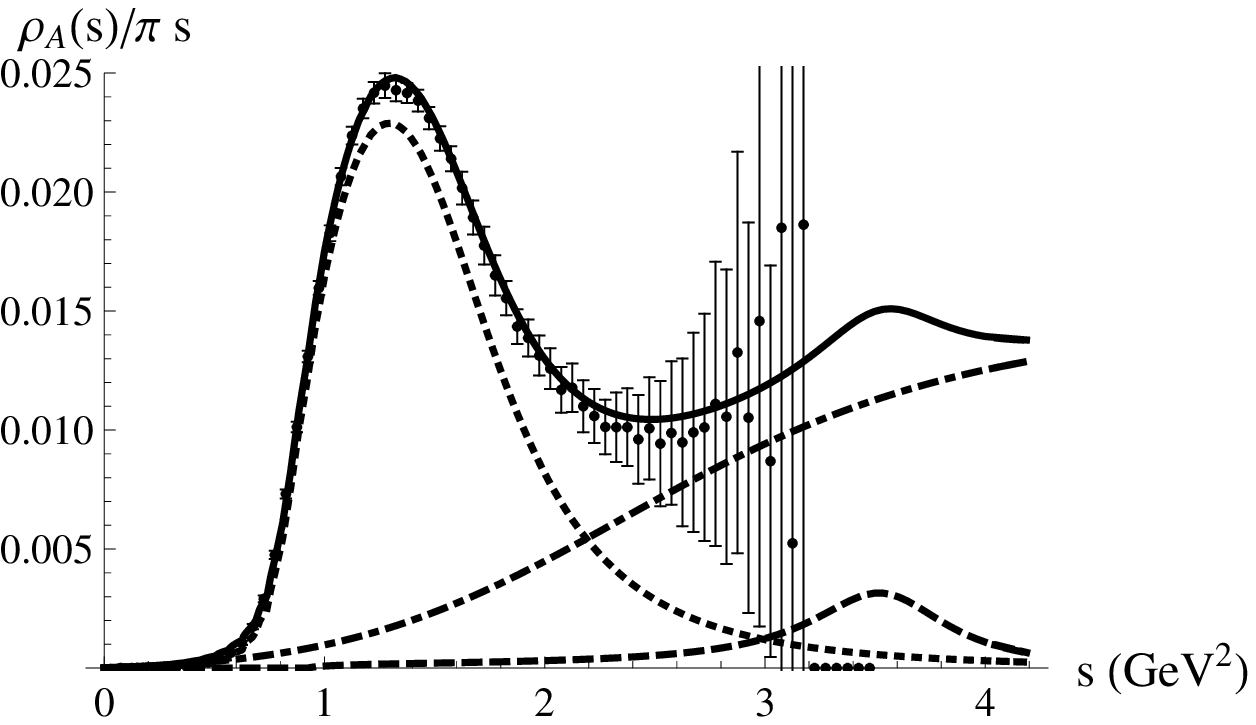}
\caption{Vacuum spectral functions in the vector (top) and axialvector (bottom) channels,
compared to experimental data for hadronic $\tau$ decays~\cite{Barate:1998uf};
The total spectral function in each channel (solid curve) is composed of a
ground state (dotted curve), excited resonance (dashed
curve), and a universal continuum (dot-dashed curve).
}
\label{fig:vacsf}
\end{figure}

Our starting point are the vacuum axial-/vector spectral functions of
Ref.~\cite{Hohler:2012xd}\footnote{The normalization used in Eq.~(25) of
Ref.~\cite{Hohler:2012xd} for the Breit-Wigner width of the $a_1$ peak contained a (small)
imaginary contribution; we have corrected this and could recover the same level of agreement
with the experimental data and sum rules with a minor modification of the parameters.}.
They are comprised of contributions from the ground state ($\rho$ and $a_1$ peaks), a first
excited state ($\rho'$ and $a_1'$), and a chirally invariant (i.e., identical) continuum
for both channels. The vacuum $\rho$ is taken from the microscopic model of
Ref.~\cite{Urban:1998eg}, while $a_1$, $\rho'$ and $a_1'$ are parameterized with Breit-Wigner
functions. For the present analysis, we have slightly modified the vacuum parameters of the
$\rho'$ to shift its threshold energy to higher energies. This avoids its low-mass tail to
reach well below 1\,GeV where the $\tau$-decay data do not exhibit any 4$\pi$ contributions.
The  modification to the $\rho'$ formfactor is compensated by a small modification of the
mass and width of the $a_1'$ as to recover a near-perfect agreement with WSR-1 and WSR-2. The
re-evaluation of the vacuum QCDSRs requires numerical values of 4-quark factorization
parameter of $\kappa$=2.1 in $\avg{\mathcal{O}_4^{SB}}=\frac{16}{9}\kappa \avg{\bar qq}^2$,
and of the gluon condensate of
$\avg{\frac{\alpha_s}{\pi} G^2_{\mu\nu}}$=0.017\,GeV$^4$.
The updated vacuum spectral functions, shown
in Fig.~\ref{fig:vacsf}, are very similar to the ones in Ref.~\cite{Hohler:2012xd}.


Finite-temperature effects in the spectral functions are implemented as follows.
For the $\rho$ meson, we employ the microscopic calculations using hadronic effective
theory~\cite{Rapp:1999us} at vanishing baryon chemical potential. This is the key input to
our analysis, as these spectral functions are consistent with dilepton data in
URHICs~\cite{Rapp:2013ema}, and thus provide a direct link to experiment. The only amendment
we allow is a reduction of the vector-dominance coupling strength (as routinely done in
QCDSR analyses~\cite{Hatsuda:1992bv,Zschocke:2002mn,Leupold:1997dg,Leupold:2001hj}). Optimal
agreement with the QCDSR requires a reduction of up to 7\% at $T$=170\,MeV.

For the $a_1$ meson, the lack of quantitative calculations at finite $T$ leads us to
parameterize the medium modifications of its spectral function. We introduce four
parameters which control the $a_1$ peak's location, width, and strength in-medium. For
the $a_1$ mass, we write $M^T_{a_1} = M_{a_1} (1-\delta M_{a_1}(T)/M_{a_1})$,
and for the current coupling
$C^T_{a_1} = C_{a_1} (1-\delta C_{a_1}(T)/C_{a_1})$.
The width is increased and extended below the vacuum threshold by adding the
following term to the vacuum width, $\Gamma_{a_1}(s)$,
\begin{equation}
\Delta \Gamma_{a_1}(s) = \left(\Gamma_1^T + \frac{s}{M_{a_1}^2} \Gamma_2^T \right)
\left(\frac{\Lambda_{a_1}^2 + M_{a_1}^2}{\Lambda_{a_1}^2+s}\right)^2
\end{equation}
where $\Gamma_1^T$ and $\Gamma_2^T$ are $T$-dependent constants, and the last factor is a
formfactor with the same scale, $\Lambda_{a_1}$, as in vacuum.
The resulting  ground-state axialvector spectral function in medium takes the form
\begin{equation}
\rho_{a_1}(s, T) = \frac{1}{\pi} {C}^T_{a_1} \frac{\sqrt{s} \,
\Gamma_{a_1}^T(s,T)}{(s-M_{a_1}^{T 2})^{2} + s \Gamma_{a_1}^T(s,T)^2} \ ,
\end{equation}
with $\Gamma_{a_1}^T(s,T) = \Gamma_{a_1}(s) + \Delta \Gamma_{a_1}(s)$.

The temperature dependence of the excited states is even less known. Instead of introducing
additional parameters for their in-medium Breit-Wigners (which are hard to control), we
rather apply the model
independent low-temperature effect known as chiral mixing~\cite{Dey:1990ba,Steele:1996su}
to the $\rho'$ and $a_1'$ states. However, in the spirit of the HRG, we go beyond the mixing
induced by only thermal pions by including the effect from the virtual pion cloud of the
thermal hadrons.
This effect has been worked out for the pion cloud of the nucleon in cold nuclear
matter~\cite{Chanfray:1998hr,Krippa:1997ss}.
To extend it to other hadrons (not including the non-pion Goldstone bosons),
we define a mixing parameter
\begin{equation}
\hat{\epsilon}_h (T) = \frac{4}{3}\frac{\sigma^{\rm cloud}_\pi}{f_\pi^2 m_\pi^2}\varrho_s^h \ .
\end{equation}
The total mixing parameter, $\hat{\epsilon}$, is the sum of the individual
$\hat{\epsilon}_h$ plus that of the pion, $\hat{\epsilon}_\pi =2\varrho_s^\pi/(3m_\pi f_\pi^2)$.
As with the quark condensates, we introduce an additional $T^{10}$-term to render
$\hat{\epsilon}=1/2$ at the temperature where $\avg{\bar{q}q}_T = 0$.
The in-medium spectral functions for the excited  axial-/vector states then follow as
\begin{equation}
\begin{split}
&\rho_{V'}(T) = [1-\hat{\epsilon}(T)]\rho_{V'}^{\rm vac}+\hat{\epsilon}(T) \,
\rho_{A'}^{\rm vac} + \frac{1}{2}\, \hat{\epsilon}(T) \,\rho_{a_1}^{\rm vac} \, , \\
&\rho_{A'}(T) = [1-\hat{\epsilon}(T)] \rho_{A'}^{\rm vac} + \hat{\epsilon}(T) \rho_{V'}^{\rm vac}\, .
\end{split}
\end{equation}
The $a_1$ contribution to the excited vector channel admixes only the part
which is not included in the microscopic calculation of the $\rho$, see
Ref.~\cite{vanHees:2007th} for details.
Our approximate extension of the mixing beyond the low-$T$ pion gas limit is only carried
linear in the (scalar) hadron densities, but in line with the in-medium treatment
of the condensates. However, no finite-momentum nor finite-mass effects of the (virtual)
pions have been accounted for.

The chirally invariant continuum is assumed to be $T$-independent (e.g., chiral
mixing would not affect it).

Lastly, we need to address the $T$ dependence of the 4D longitudinal part of the
axial-vector spectral function, i.e., the pion pole. We approximate the pion mass
by the leading-order prediction of chiral perturbation theory,
\begin{equation}
m_\pi^2(T) = m_\pi^2\left(1+\frac{1}{4}\hat{\epsilon}_\pi(T)\right),
\end{equation}
{\it i.e.}, induced by the pion gas only. This produces
a weak $T$ dependence as expected for a Goldstone boson. Assuming the
Gell-Mann--Oakes--Renner relation to hold at finite $T$,  allows us to infer $f_\pi(T)$
from the above-constructed $T$-dependence of the quark condensate.

To summarize this section, we have supplemented a microscopic model for the $\rho$ spectral
function with a 4-parameter ansatz for the in-medium $a_1$, chiral mixing for the excited
states, and a weakly $T$-dependent pion mass from chiral perturbation theory.
We now investigate whether this setup can satisfy QCDSRs and WSRs.

\section{Finite-Temperature Sum Rule Analysis}
\label{sec:results}

\begin{figure*}[!t]
\centering
\includegraphics[width=.9\textwidth]{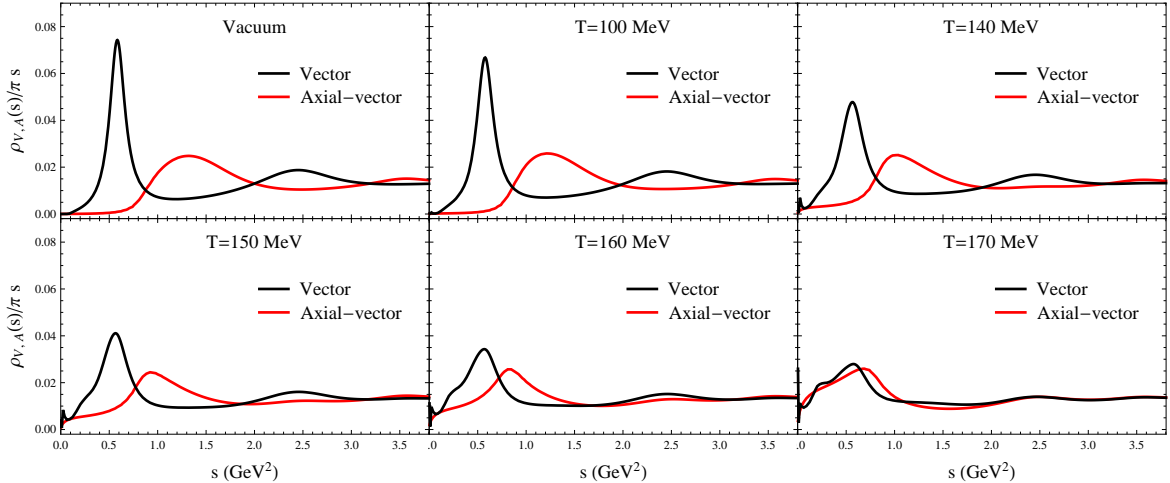}
\caption{Finite-temperature vector (black curve) and axialvector (red curve) spectral functions.}
\label{fig:sf}
\end{figure*}

Let us start by describing the quantitative criteria which govern the numerical
values of the in-medium $a_1$ parameters introduced in the previous section.
\begin{table}[!b]
\begin{center}
\begin{tabular}{|c|cccccc|}
\hline
$T$ [MeV] & 0  & 100 & 140 & 150 & 160 & 170 \\
\hline
$d_V (\%)$ & 0.59 & 0.43 & 0.44 & 0.49 & 0.57 & 0.67 \\
$d_A (\%)$ & 0.49 & 0.48 & 0.56 & 0.59 & 0.55 & 0.56 \\
\hline
$d_{\rm WSR1} (\%)$& $\sim 0$ & 0.003 & 0.04 & 0.04 & -0.004 & 0.004 \\
$d_{\rm WSR2} (\%)$& $\sim 0$ & -0.0002 & -0.0008 & -0.002 & -0.0003 & -0.005\\
$d_{\rm WSR3} (\%)$& 200 & 181 & 258 & 372 & 585 & 11600\\
\hline
$r_{-1}$ & 1 & 0.96 & 0.72 & 0.57 & 0.37 & 0.14\\
$r_{0}$ & 1 & 0.93 & 0.66 & 0.50 & 0.31 & 0.12\\
$r_{1}$ & 1 & 0.91 & 0.64 & 0.50 & 0.32 & 0.15\\
\hline
\end{tabular}
\end{center}
\caption{Summary of deviation measures for QCDSRs (upper 2 lines) and WSRs (lower 6 lines)
at finite temperature.}
\label{tab:results}
\end{table}

To evaluate the QCDSRs, we adopt the conventional method of
Refs.~\cite{Leinweber:1995fn,Leupold:1997dg} to calculate an average deviation between the
LHS and RHS over a suitable Borel window, referred to as a $d$-value. The same procedure and
Borel window criteria as for the vacuum analysis in Ref.~\cite{Hohler:2012xd} are adopted.
A $d$-value of below 1\% has been argued to reasonably bracket remaining uncertainties in the
matching procedure~\cite{Leupold:1997dg}; we adopt this as our figure of merit in both $A$ and
$V$ channels below.

To evaluate the WSRs, we define a similar measure of deviation between the two sides as
\begin{equation}
d_{\rm WSR} = \frac{{\rm LHS} - {\rm RHS}}{{\rm RHS}} \ .
\end{equation}
This measure is much simpler than the QCDSR analog because it does not involve any Borel
window. However, it also has its subtleties. The integrands of the LHS of each WSR are
oscillatory functions with appreciable cancelations to yield the RHS (cf.~Fig.~2 in
Ref.~\cite{Hohler:2012xd}), especially for the higher moments. Since we only use a finite
number of moments (3), this could, in principle, lead to ``fine-tuned solutions" to the
WSRs where the oscillations are still large, and thus $\rho_V(s)\ne \rho_A(s)$ even close
to restoration. To probe this behavior (and thus the sensitivity to any ``artificial"
fine tuning), we introduce an ``absolute-value" version of the LHS by
\begin{equation}
\tilde{w}_n(T) \equiv \int_0^\infty ds \ s^{n} \ |\Delta \rho(s;T)| \ .
\end{equation}
Though these moments are not directly related to chiral order parameters, they
should diminish toward restoration. We define pertinent ratios
$r_n = \tilde{w}_n(T)/\tilde{w}_n(T=0)$.

Our analysis proceeds as follows. We first evaluate the QCDSR for the vector channel.
With a small reduction in the vector dominance coupling, we find acceptable $d_V$ values
ranging from 0.43\% to 0.67\% for all $T$=0-170\,MeV (cf.~Tab~\ref{tab:results}).
This is a nontrivial result by itself.
For the axialvector channel, the QCDSRs and two WSRs are used simultaneously to search
for in-medium $a_1$ parameters which minimize
\begin{equation}
f = d_{\rm WSR1}^2 + d_{\rm WSR2}^2  + d_A^2 \ ,
\label{f}
\end{equation}
while requiring a smooth $T$ dependence.
The thus obtained finite-$T$ axialvector spectral functions are shown
in Fig.~\ref{fig:sf}. For all cases, the percentage deviation of WSR-1 and WSR-2 is below
0.1\%, and $d_A$ remains below 0.6\%. Deviations of WSR-3 are much
larger, but comparable to the vacuum up to $T$$\simeq$150\,MeV. At $T$=160 and
especially 170\,MeV, the magnitude of the RHSs is small and enters into the denominator
of $d_{\rm WSR}$, thus greatly magnifying residual deviations. The $r_n$ measures
decrease monotonically with $T$ suggesting acceptable deviations even for WSR-3.
We therefore conclude that our spectral functions are compatible with both
QCDSRs and WSRs.

\begin{figure}[!t]
\centering
\includegraphics[width=.45\textwidth]{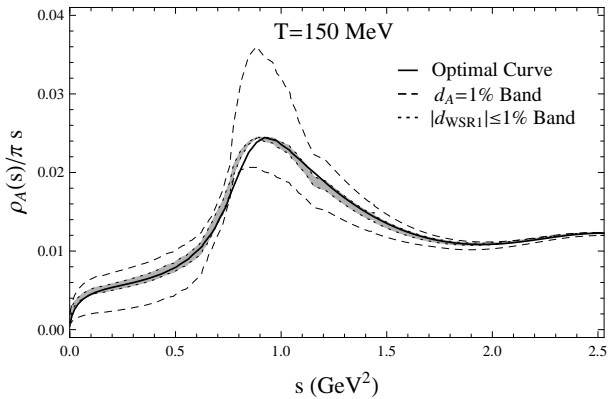}
\caption{Regions of axialvector spectral functions at $T$=150\,MeV
when requiring agreement with the QCDSR only at $d_A$=1\% (dashed lines),
and additionally with WSR-1 at $|d_{\rm WSR1}|$$\leq$1\% (dotted lines).
The solid line corresponds to a minimal $f$ value from Eq.~(\ref{f}).}
\label{fig:axialband}
\end{figure}

To probe the uncertainties in our method,
we depict in Fig.~\ref{fig:axialband} ranges of axialvector spectral functions with
relaxed constraints, at an intermediate temperature of $T$=150\,MeV.
The dashed lines border a regime of spectral functions which are obtained by only
requiring $d_A$=1\% for the axialvector QCDSR (the band could be larger if all
spectral functions with $d_A$$<$1\% were included). From this collection of curves,
we then select those whose agreement with WSR1 is within $1\%$, producing a much
narrower (shaded) region bordered by dotted lines. The {\em combined} constraints of
QCDSRs and WSRs are thus shown to noticeably increase the selectivity of the in-medium
axialvector spectral function.

A visual inspection of the in-medium spectral functions supports the trend toward restoration,
cf.~Fig.~\ref{fig:sf}: the $a_1$ peak gradually merges into the $\rho$ while the excited
states degenerate somewhat earlier through chiral mixing. The $\rho$-$a_1$ merging is
largely dictated by the WSRs, but the concrete shape close to chiral restoration is more
sensitive to the QCDSRs. Note that our analysis not only complies with a ``trivial"
degeneracy at the restoration point, but rather provides a systematic
temperature evolution, starting from the vacuum, compatible with current best
estimates for the $T$ dependent chiral order parameters and condensates (at $T$=170\,MeV,
our condensates are close to zero, undershooting the lQCD data for the 2-quark condensate;
our axialvector spectral function at this temperature is thus more of an illustration of the
expected degeneracy at higher $T$ where $\avg{\bar qq}_T$$\simeq$0).
The in-medium $a_1$ mass shift is consistent with a leading $T^4$ behavior, in line
with model-independent constraints from the chiral Lagrangian.
Our analysis also suggests that the approach toward restoration ``burns off" the chiral
mass splitting between the
$\rho$ and $a_1$, while ``bare" masses of $m_0$$\simeq$0.8\,GeV essentially
persist, similar to Ref.~\cite{Urban:2001uv}.

\section{Conclusion}
\label{sec:conc}
The objective of this work was to test whether in-medium vector spectral functions
which describe dilepton data in heavy-ion collisions are compatible with chiral
symmetry restoration. Toward this end, we deployed QCD and Weinberg sum rules in a
combined analysis of vector and axialvector spectral functions, using lattice-QCD
and the hadron resonance gas to estimate the in-medium condensates and chiral order
parameters, and chiral mixing to treat the $T$ dependence of excited states. We first
found that the QCDSR in the vector channel is satisfied with a small (order 5\%)
amendment of vector dominance. We then introduced a 4-parameter ansatz for the
in-medium $a_1$ spectral function and found that a smooth reduction of its mass
(approaching the $\rho$ mass) and large increase in width (accompanied by a low-mass
shoulder) can satisfy the axialvector QCDSR and 3 WSRs over the entire temperature
range from $T$=0-170\,MeV, ultimately merging with the vector channel.
This establishes a direct connection between dileptons and chiral restoration, and thus
the answer to the originally raised question is positive. Our findings remain to be
scrutinized by microscopic calculations of the $a_1$ spectral function. Work in this
direction is ongoing.

\acknowledgments
This work is supported by the US-NSF under grant No.~PHY-1306359 and
by the A.-v.-Humboldt Foundation (Germany).


\end{document}